# Nature of polarization fatigue in BiFeO$_3$ **

*By Seung-Hyub Baek, Chad M. Folkman, Jae-Wan Park, Sanghan Lee, Chung-Wung Bark, Thomas Tybell and Chang-Beom Eom\**

[*]    Prof. C. B. Eom, S. H. Baek, C. M. Folkman, J. W. Park, S. Lee, C. W. Bark
Department of Materials Science and Engineering,
University of Wisconsin, Madison, WI 53706 (USA)
E-mail: eom@engr.wisc.edu

       Prof. Thomas Tybell
Department of Electronics and Telecommunications,
Norwegian University of Science and Technology, 7491 Trondheim, Norway

Department of Materials Science and Engineering,
University of Wisconsin, Madison, WI 53706 (USA)

[**] This work is supported by the Army Research Office under Grant No. W911NF-10-1-0362

Keywords: BiFeO$_3$, fatigue, switching, orientation, ferroelectric

As a room-temperature multiferroic, BiFeO$_3$ has been intensively investigated for both magnetoelectric devices and non-volatile ferroelectric memory applications.[1-3] BiFeO$_3$, having a rhombohedral unit cell, has antiferromagnetic, ferroelectric and ferroelastic order parameters. Since BiFeO$_3$ exhibits coupling between spontaneous electric polarization in the [111] direction and the (111) antiferromagnetic planes, control of the magnetic order can be achieved by ferroelectric domain reorientation resulting from polarization switching by an external electric field.[4] This possibility of controlling the magnetism by an electric field has been demonstrated at room temperature in single crystals[5, 6] and thin films[4, 7]. In addition, BiFeO$_3$ has the largest remanent polarization ($P_r$ ~100 μC cm$^{-2}$) along the [111] polar direction among all known ferroelectrics,[1-3] which is a promising feature as a lead-free material for ferroelectric random access memory (FeRAM). Utilizing the large remanent polarization of BiFeO$_3$ would enable further reduction of the cell size limited by conventional ferroelectrics such as BaTiO$_3$ and Pb(Zr,Ti)O$_3$.

Both magnetoelectric and ferroelectric memory devices have the same control knob: polarization switching by an applied electric field.[1-7] Due to the rhombohedral symmetry of BiFeO$_3$, there are four ferroelastic variances and three different polarization switching events: (1) 71° switching from r1$^-$ to r3$^+$, (2) 109° switching from r1$^-$ to r2$^+$ (or r4$^+$), and (3) 180° switching from r1$^-$ to r1$^+$ (the superscript + and - stand for up and down polarization, respectively). Each switching path is coupled to a different reorientation of the BiFeO$_3$ unit cell, and hence different coupling to the magnetic order[4] as well as different magnitudes of switchable polarization.[8, 9] A degradation of the ferroelectric properties of BiFeO$_3$ will result in losing controllability of magnetic order switching in magnetoelectric devices and capacity for information storage in ferroelectric memory devices. Especially, polarization fatigue[10, 11] will directly restrict the reliability of the actual devices. Hence it is important to understand the intrinsic fatigue behavior of each polarization switching path in BiFeO$_3$ thin films. In this communication, we report polarization fatigue in BiFeO$_3$ depending on switching path, and propose a fatigue model which will broaden our understanding of the fatigue phenomenon in low-symmetry materials.

Previously, there were reports on fatigue characteristics of rhombohedral ferroelectrics: *ferroelastic domain evolution* with polarization fatigue in textured films[12, 13] and ceramics[14, 15] of PZN-PT, and *orientation dependence* showing (111)-oriented samples are more easily fatigued than (001)-oriented ones in PZN-PT[16-18] and BiFeO$_3$[19]. In order to study the intrinsic behavior of switching-path dependent fatigue, it is crucial (1) to control a single polarization switching path among the three possible ones (71°, 109° and 180°) during switching cycles[20, 21] and (2) to remove the extrinsic effects of the pre-existing domain or grain boundaries affecting polarization switching.[9, 22-24] To solve the latter issue, we used *monodomain* epitaxial BiFeO$_3$ thin films as a model system, without the extrinsic effects of pre-existing domain walls.[21] All three orientations of monodomain BiFeO$_3$ films in this work have a downward polarization direction, hence r1$^-$ domain. In order to achieve the former requirement, we used *three different crystallographic orientations*, (001)$_{pc}$ for 71°, (110)$_{pc}$ for 109° and (111)$_{pc}$ for 180° switching, of epitaxial monodomain BiFeO$_3$ films in the vertical capacitor structure of Pt top and SrRuO$_3$ bottom electrodes (the subscript "pc" stands for pseudocubic). We used a large size of top electrodes (>50 μm) to prevent ferroelastic back-switching for a reliable switching path control.[21] Combined XRD measurements and initial switching experiments revealed that all three switching paths were fully reversible, see supporting information.

In order to elucidate how the switching path affects the fatigue behavior, 10-μs-wide pulses with a repetition frequency of 200 Hz were applied to each orientation film through the Pt top electrode. **Figure** 1a-c show *P-E* measurements before and after the fatigue cycling of the (001)$_{pc}$, (110)$_{pc}$ and (111)$_{pc}$ BiFeO$_3$ films, respectively. Figure 1d depicts the switched polarization versus the number of switching cycles. Initial ~115 μC cm$^{-2}$ of $P_r$ in mono-domain (111)$_{pc}$ BiFeO$_3$ film starts to degrade at ~10$^4$ cycles, and reduces to ~57 μC cm$^{-2}$ at $10^6$ cycles. On the contrary, $P_r$ of mono-domain $(001)_{pc}$ and $(110)_{pc}$ BiFeO$_3$ films continues unabated even up to $10^6$ cycles, the maximum number of cycles in this study.

We performed macroscopic analysis on $(111)_{pc}$ BiFeO$_3$ films. **Figure** 2a and 2b show the RSM around the 113 SrTiO$_3$ peak before and after fatigue cycles were applied to a $(111)_{pc}$ monodomain BiFeO$_3$ film, respectively. The additional peaks for BiFeO$_3$ indicate that new ferroelastic domains formed during fatigue cycles and that fatigued $(111)_{pc}$ BiFeO$_3$ films have four ferroelastic variances. It should be noted that these new ferroelastic domains are *nucleated from the initial monodomain state* with fatigue. We also investigated the microscopic domain structure using PFM to understand the configuration of new ferroelastic domains as well as to obtain local information related to electrical data of $(111)_{pc}$ BiFeO$_3$ films. We applied $10^5$ switching cycles on $(111)_{pc}$ BiFeO$_3$ film with the final polarity pointing upward. The Pt top electrode was subsequently removed by ultrasonification, and the domain structure analyzed with PFM. The as-grown $(111)_{pc}$ monodomain BiFeO$_3$ film does not show any contrasts in out-of-plane (OP) and in-plane (IP) PFM images (figure 2c, d). However, PFM images of fatigued $(111)_{pc}$ BiFeO$_3$ film show contrasts coming from new domains in figure 2e (OP) and 2f (IP). The OP image has three different contrasts: dark, grey and bright. This indicates the OP polarization component has three different states, see supporting information. Considering both OP and IP images, we confirmed that new ferroelastic (ferroelectric) domains are formed such as $r2^{\pm}$, $r3^{\pm}$ and $r4^{\pm}$ rather than $r1^{\pm}$ during the fatigue cycles, consistent with the XRD data in figure 2b. On the other hand, the $(001)_{pc}$ and $(110)_{pc}$ monodomain BiFeO$_3$ films do not show any PFM contrasts from new domains. Moreover, the reduction in remanent polarization for the fatigued $(111)_{pc}$ BiFeO$_3$ is consistent with the area of newly formed ferroelastic domains:

$$\frac{\text{Pr (fatigued)}}{\text{Pr (initial)}} \approx \frac{\text{Area (new ferroelastic domains)}}{\text{Area (total measured)}}$$ within ±10 % error as measured by PFM on

five different areas of 5 μm × 5 μm within a fatigued capacitor. Hence, it is concluded that polarization fatigue of $(111)_{pc}$ BiFeO$_3$ is directly related to the formation of new domains.

In order to elucidate the fatigue mechanism in $(111)_{pc}$ BiFeO$_3$ films, we analyzed the configuration of the domains and the domain walls in detail. We identified non-neutral domain walls with a head-to-head configuration in the fatigued $(111)_{pc}$ BiFeO$_3$ films. Figure 2g is an IP PFM image of the area marked as a blue dotted square in figure 2e and 2f. Green and grey dotted lines correspond to the bright and grey contrasts in the OP image (figure 2e), respectively. Non-neutral domain walls are marked as sky-blue (between positive intermediate domains such as r2$^+$, r3$^+$ and r4$^+$) and pink solid lines (between negative intermediate domains such as r2$^-$, r3$^-$ and r4$^-$). Most of the domain walls are meandering without preferential crystallographic alignments, indicating that the domains and domain walls are not a result of reducing mechanical and electric energy, related to neutral and aligned domain walls,[22, 23, 25, 26] but that extrinsic factors (defects) are important. This agrees with a domain wall pinning mechanism, one of the scenarios widely believed for fatigue, where electrostatic coupling between a non-neutral domain wall and mobile carriers forms a stable (pinned) electro-neutral structure.[27, 28]

In order for non-neutral domain boundaries to be formed as a domain wall pinning center, two conditions have to be satisfied: (1) the formation of a head-to-head (or tail-to-tail) polarization configuration during switching should be possible, and (2) mobile carriers should be present. The latter are common oxide thin films, such as oxygen vacancies, even though they might be varied according to film deposition conditions. We believe that the concentration of mobile carriers was not varied at large between three orientations because all the BiFeO$_3$ thin films were grown using the same growth conditions. Moreover, we have found that a unipolar electric stress with the same profile as in figure 1 does not induce fatigue in $(111)_{pc}$ BiFeO$_3$ films even after $10^6$ cycles. That is, the fatigue behavior is closely related to the polarization switching event. Thus, the polarization switching path would be closely related to the degree of chance of forming a non-neutral domain walls. In this regard, we adopted the concept of the *multi-step process of 180° polarization switching*, where 180° switching consists of intermediate ferroelastic switchings, predicted by Kubel and Schmid,[29] and in agreement with relaxation-mediated 180° switching observed in BiFeO$_3$ thin films.[21] This multi-step switching process has been supported by acoustic measurements,[30] neutron diffraction,[31] PFM[32] and direct elastic measurement[33] using rhombohedral PZN-PT and PMN-PT single crystals.

Based on the multi-step switching process, 180° switching can have a maximum of six different routes from r1$^-$ to r1$^+$ through three times two of 71° switchings as shown in **Figure 3a**. When switched back, polarization will also go through the same intermediate ferroelastic states. Due to the 3-fold symmetry of (111)$_{pc}$ BiFeO$_3$ films with a [111]$_{pc}$ axis, there is no preferential path among the six. Therefore, polarization switching will follow any of the three possible paths *with equal probability* across the sample. Thus, there is a large chance of forming non-neutral domain walls when two different intermediate paths meet. Such non-neutral domain walls are expected to disappear rapidly since they are not stable. However, they can be stabilized by trapping mobile charge carriers to reduce high electrical energy at domain walls, as shown in figure 3b. This stabilized structure would pin the domain walls, leading to fatigue. In contrast, 71° switching has only one path from r1$^-$ to r3$^+$, as shown in figure 3c. During switching cycles, the switched and the unswitched regions maintain neutral domain walls with head-to-tail configurations. Thus, the chance of forming non-neutral domain walls is negligible during 71° switching cycles in (001)$_{pc}$ films, resulting in fatigue resistance. Fatigue-resistant 109° switching can be also understood from this perspective. This model implies that the fatigue of the (111)$_{pc}$ BiFeO$_3$ film is a *stochastic* event happening inside the film rather than an interfacial effect. The nearly equal population of six ferroelectric domains in figure 3 also indirectly indicates the probabilistic character of fatigue. This result suggests that multi-step 180° switching in (111)$_{pc}$ BiFeO$_3$ films is not favorable for a real device in terms of the reliability issue even though this has the largest remanent polarization.

In summary, we studied intrinsic polarization fatigue depending on the switching path using *monodomain* epitaxial BiFeO$_3$ thin films. We controlled each switching path selectively by using different orientations of BiFeO$_3$ films. The 180° switching path in (111)$_{pc}$ films turned out to be vulnerable to fatigue while 71° switching in (001)$_{pc}$ and 109° switching in (110)$_{pc}$ were fatigue-resistant. Our microscopic analysis with PFM showed direct evidence of the pinned domain walls with non-neutral configuration of polarization, which is consistent with macroscopic analysis of polarization fatigue using XRD and electrical measurements. We proposed a model that the complex multi-step switching process of 180° polarization reversal results in domain wall pinning by incorporation of mobile charge carriers into non-neutral domain walls. This work provides a framework for understanding electric fatigue in BiFeO$_3$, and other low symmetric materials with complex switching routes. In addition, this work provides design rules for the reliable performance of multifunctional devices controlled by polarization switching.

**Experimental**

Mono-domain BiFeO$_3$ films with (001)$_{pc}$, (110)$_{pc}$ and (111)$_{pc}$ orientations were grown with the off-axis sputtering method on (001) SrTiO$_3$ substrates miscut by 4° along [100], (110) SrTiO$_3$ substrates miscut by 0.3° along [1-11], and (111) SrTiO$_3$ substrates, respectively. All BiFeO$_3$ films were grown in the same growth conditions.[9] As a bottom electrode, 50-nm SrRuO$_3$ with smooth surface[34] was deposited on each orientation of SrTiO$_3$ substrates by 90° off-axis sputtering before BiFeO$_3$ was deposited. Pt top electrodes were formed by rf sputtering and photolithography. A high-resolution four-circle x-ray diffraction, a D8 Advance (Bruker AXS), was used for HRXRD-RSM measurements. A Radiant PFH100 ferroelectric measurement system was used to apply electric pulse stress for fatigue measurement and measure the *P-E* hysteresis loop. The ferroelectric domain was investigated using a DI Multimode AFM system with a Nanoscope V controller. Pt/Ir- and Cr-coated conductive tips were used.

# References

[1] H. Bea, M. Gajek, M. Bibes and A. Barthlelmy, *J. Phys.: Condens. Matter* **2008**, 20, 434221.

[2] G. Catalan, J. F. Scott, *Adv. Mater.* **2009**, 21, 2463.

[3] L. W. Martin, S. P. Crane, Y. H. Chu, M. B. Holcomb, M. Gajek, M. Huijben, C. H. Yang, N. Balke, R. Ramesh, *J. Phys.: Condens. Matter* **2008**, 20, 434220

[4] T. Zhao, A. Scholl, F. Zavaliche, K. Lee, M. Barry, A. Doran, M. P. Cruz, Y. H. Chu, C. Ederer, N. A. Spaldin, R. R. Das, D. M. Kim, S. H. Baek, C. B. Eom, R. Ramesh, *Nature Mater.* **2006**, *5*, 823.

[5] D. Lebeugle, D. Colson, A. Forget, M. Viret, A. M. Bataille, A. Gukasov, *Phys. Rev. Lett.* **2008**, 100, 227602.

[6] S. Lee, W. Ratcliff II, S-. W. Cheong, and V. Kiryukhin1, *Appl. Phys. Lett.* **2008**, 92, 192906.

[7] Y. H. Chu, L. W. Martin, M. B. Holcomb, M. Gajek, S. J. Han, Q. He, N. Balke, C. H. Yang, D. Lee, W. Hu, Q. Zhan, P. L. Yang, A. Fraile-Rodriguez, A. Scholl, S. X. Wang, R. Ramesh, *Nature Mater.* **2008**, 7, 478.

[8] J. F. Li, J. L. Wang, M. Wuttig, R. Ramesh, N. Wang, B. Ruette, A. P. Pyatakov, A. K. Zvezdin, D. Viehland, *Appl. Phys. Lett.* **2004**, 84, 5261.

[9] R. R. Das, D. M. Kim, S. H. Baek, C. B. Eom, F. Zavaliche, S. Y. Yang, R. Ramesh, Y. B. Chen, X. Q. Pan, X. Ke, M. S. Rzchowski, S. K. Streiffer, *Appl. Phys. Lett.* **2006**, 88, 242904.

[10] A. K. Tagantsev, I. Stolichnov, E. L. Colla, and N. Setter, *J. Appl. Phys.* **2001**, 90, 1387.

[11] X. J. Lou, *J. Appl. Phys.* **2009**, 105, 024101.

[12] J. S. Liu, S. R. Zhang, L. S. Dai, Y. Yuan, *J. Phys. Lett.* **2005**, 97, 104102.

[13] V. V. Shvartsman, A. L. Kholkin, C. Verdier, D. C. Lupascu, *J. Phys. Lett.* **2005**, 98, 094109.

[14] N. Menou, Ch. Mullera, I. S. Baturin, V. Ya. Shur, J. L. Hodeau, *J. Phys. Lett.* **2005**, 97, 064108.

[15] N. Zhang, L. Li, Z. Gui, *Mater. Lett.* **2002**, 56, 244.

[16] M. Ozgul, S. Trolier-Mckinstry, C. A. Randall, *J. Electroceram.* **2008**, 20, 133.

[17] V. Bornand, S. Trolier-McKinstry, K. Takemura, and C. A. Randall, *J. Appl. Phys.* **2000**, 87, 3965.

[18] K. Takemura, M. Ozgul, V. Bornand, S. Trolier-McKinstry, C. A. Randall, *J. Appl. Phys.* **2000**, 88, 7272.

[19] H. W. Jang, S. H. Baek, D. Ortiz, C. M. Folkman, C. B. Eom, Y. H. Chu, P. Shafer, R. Ramesh, V. Vaithyanathan, and D. G. Schlom, *Appl. Phys. Lett.* **2008**, 92, 062910.

[20] N. Balke, S. Choudhury, S. Jesse, M. Huijben, Y. H. Chu, A. P. Baddorf, L. Q. Chen, R. Ramesh, S. V. Kalinin, *Nature Nanotech.* **2009**, 4, 868.

[21] S. H. Baek, H. W. Jang, C. M. Folkman, Y. L. Li, B. Winchester, J. X. Zhang, Q. He, Y. H. Chu, C. T. Nelson, M. S. Rzchowski, X. Q. Pan, R. Ramesh, L. Q. Chen, C. B. Eom, *Nature Mater.* **2010**, 9, 309.

[22] H. W. Jang, D. Ortiz, S. H. Baek, C. M. Folkman, R. R. Das, P. Shafer, Y. B. Chen, C. T. Nelson, X. Q. Pan, R. Ramesh, C. B. Eom, *Adv. Mater.* **2009**, 21, 817.

[23] C. J. M. Daumont, S. Farokhipoor, A. Ferri, J. C. Wojdeł, Jorge Íñiguez, B. J. Kooi, and B. Noheda, *Phys. Rev. B* **2010**, 81, 144115.

[24] J. Seidel, L. W. Martin, Q. He, Q. Zhan, Y. -H. Chu, A. Rother, M. E. Hawkridge, P. Maksymovych, P. Yu, M. Gajek, N. Balke, S. V. Kalinin, S. Gemming, F. Wang, G. Catalan, J. F. Scott, N. A. Spaldin, J. Orenstein, R. Ramesh, *Nature Mater.* **2009**, 8, 229.

[25] S. K. Streiffer, C. B. Parker, A. E. Romanov, M. J. Lefevre, L. Zhao, J. S. Speck, *J. Appl. Phys*. **1998**, 83, 2742.

[26] C. M. Folkman, S. H. Baek, H. W. Jang, C. B. Eom, C. T. Nelson, X. Q. Pan, Y. L. Li, L. Q. Chen, A. Kumar, V. Gopalan, S. K. Streiffer, *Appl. Phys. Lett.* **2009**, 94, 251911.

[27] W. L. Warren, D. Dimos, B. A. Tuttle, G. E. Pike, R. W. Schwartz, P. J. Clews, D. C. McIntyre, *J. Appl. Phys*. **1995**, 77, 6695.

[28] J. F. Scott, M. Dawber, *Appl. Phys. Lett*. **2000**, 76, 3801.

[29] F. Kubel, H. Schmid, *Acta Cryst*. **1990**, B46, 698.

[30] W. Cao, *Ferroelectrics* **2003**, 290, 107.

[31] J. E. Daniel, T. R. Finlaysonb, M. Davis, D. Damjanovic, A. J. Studer, M. Hoffman, J. L. Jones, *J. Phys. Lett.* **2007**, 101, 104108.

[32] C. Y. Hsieh, Y. F. Chen, W. Y. Shih, Q. Zhu, W. H. Shih, *Appl. Phys. Lett*. **2009**, 94, 131101.

[33] W. Zhu, L. E Cross, *Appl. Phys. Lett*. **2004**, 84, 2388.

[34] D. Rubi, A. H. G. Vlooswijk, B. Noheda, *Thin Solid Films* **2009**, 517, 1904.

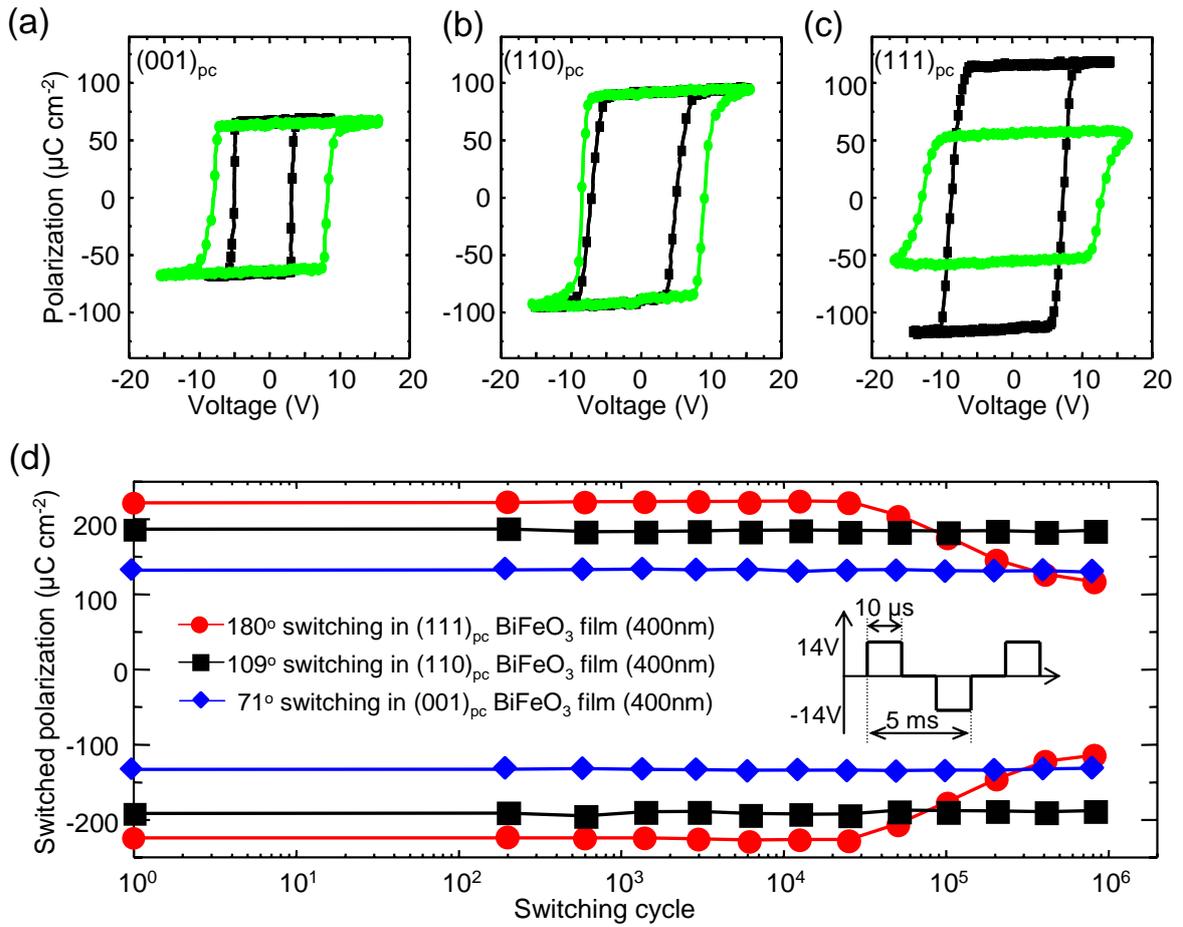

**Figure 1** *P-E* hysteresis loop measurement of initial (black square) and after fatigue cycles (green circle) of a) $(001)_{pc}$, b) $(110)_{pc}$ and c) $(111)_{pc}$ 400-nm-thick monodomain BiFeO$_3$ films with Pt top and SrRuO$_3$ bottom electrodes. d) Fatigue behavior of monodomain BiFeO$_3$ films with the three different orientations. The inset shows the electrical fatigue stress profile.

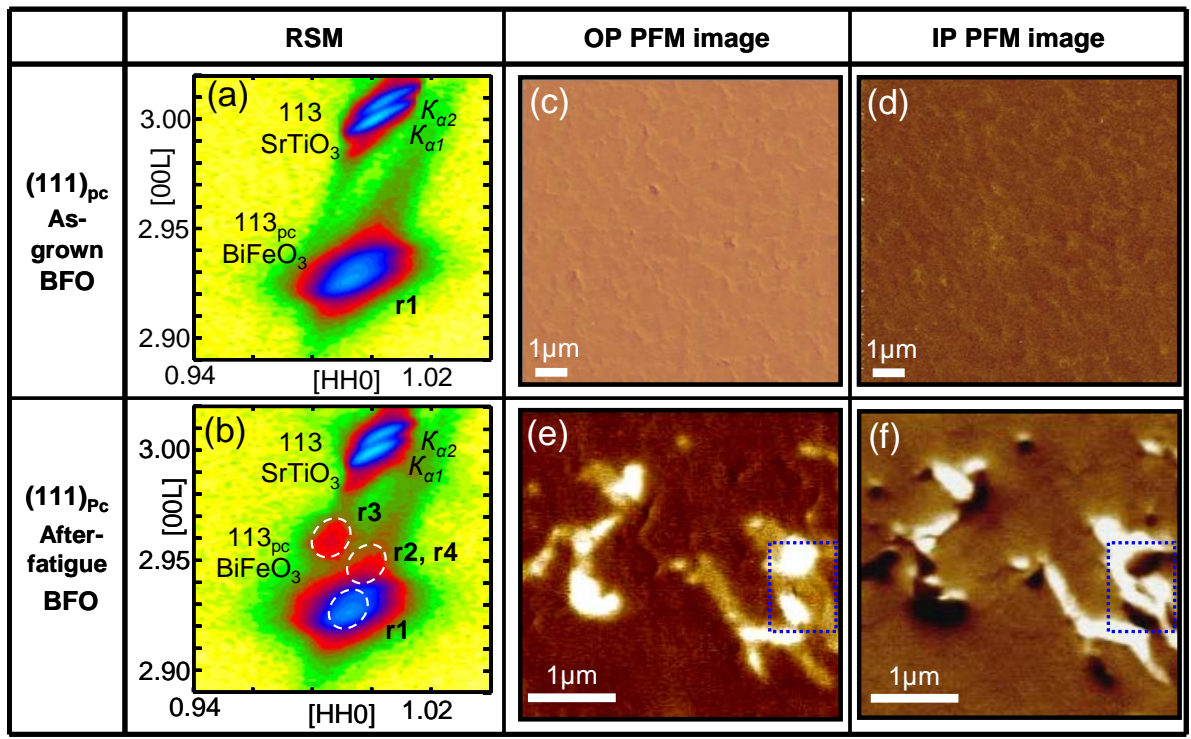
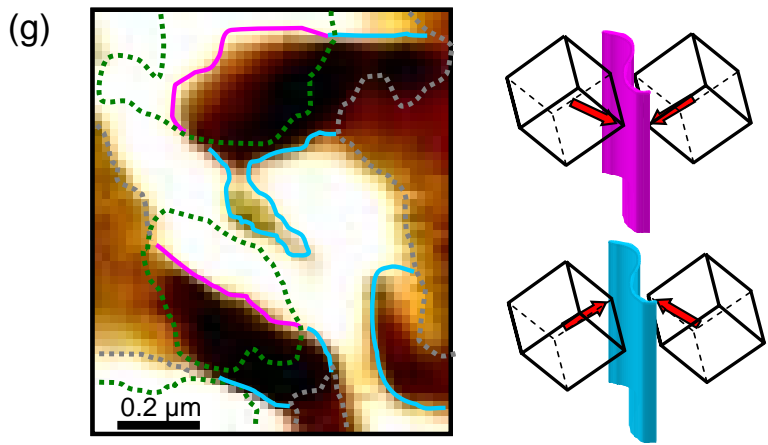

**Figure 2.** RSM data around the 113 SrTiO$_3$ peak of a) as-grown and b) fatigued (111)$_{pc}$ BiFeO$_3$ film. c) Out-of-plane and d) in-plane PFM image (phase) of the as-grown (111)$_{pc}$ BiFeO$_3$ film. e) Out-of-plane and f) in-plane PFM image (phase) of the (111)$_{pc}$ BiFeO$_3$ capacitor after 10$^5$ cycles. The phase contrast differences between white, bright-light, and dim-light in the OP image are ~10°, and ~25°, respectively. The phase contrast difference between white, light and dark in the IP image are both ~50°. g) Zoomed-in IP PFM image of the blue dotted square in e) and f). Green and grey dotted lines correspond to bright and grey contrasts in the OP image, respectively. Sky-blue and pink solid lines are non-neutral domain walls, the structures of which are also illustrated on the left, respectively.

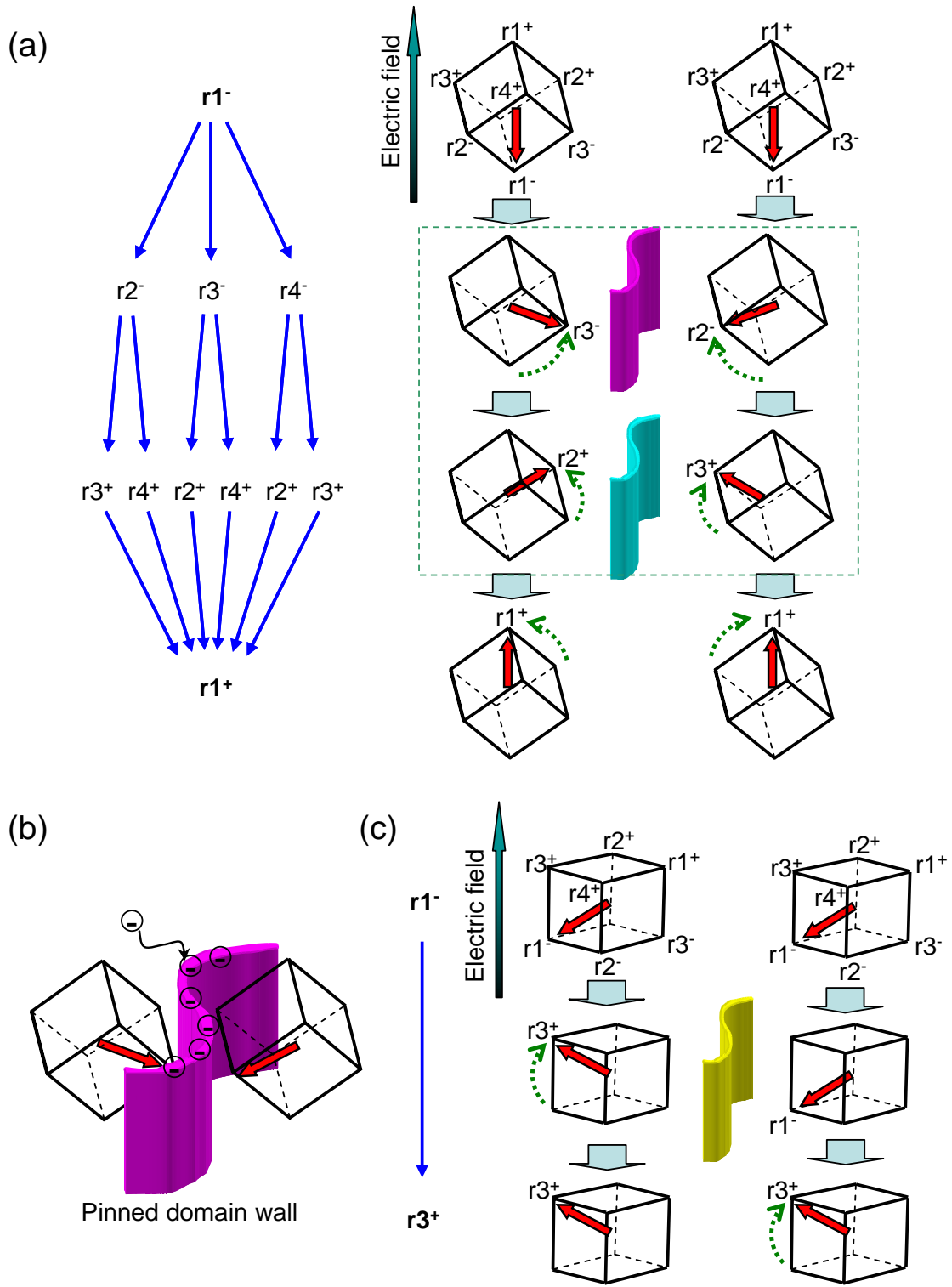

**Figure 3.** Schematic illustrations of a) complex multi-step 180° switching in (111)$_{pc}$ BiFeO$_3$ thin films and formation of temporary non-neutral domain walls, b) domain wall pinning by trapping mobile charges in the non-neutral domain walls and c) simple

71° switching in $(001)_{pc}$ BiFeO$_3$ thin films, maintaining neutral domain walls during switchings, resulting in fatigue resistance.

# Supporting information for "Nature of polarization fatigue in BiFeO$_3$"

**Polarization switching paths in BiFeO$_3$**

Due to the rhombohedral symmetry of BiFeO$_3$, there are three different polarization switching events: ferroelastic (71° and 109° polarization reversal) and ferroelectric switching (180° polarization reversal) as shown in figure S1a. (001)$_{pc}$ BiFeO$_3$ thin films have three possible switching paths under a vertical electric field: (1) 71° switching from r1$^-$ to r3$^+$, (2) 109° switching from r1$^-$ to r2$^+$ (or r4$^+$) and (3) 180° switching from r1$^-$ to r1$^+$. Because all of these paths have the same magnitude of $P_r$ projected along the [001]$_{pc}$ direction, polarization versus electric field (*P-E* hysteresis loop) measurement cannot distinguish them from each other. For the same reason, it is not possible to distinguish between 109° switching from r1$^-$ to r2$^+$ and 180° switching from r1$^-$ to r1$^+$ in (110)$_{pc}$ BiFeO$_3$ films by electrical measurement. To investigate the exact switching path of (001)$_{pc}$ BiFeO$_3$ films, we relied on a combination of electrical measurements, piezoresponse force microscopy and four circle x-ray diffraction. As shown in figure S1b-S1d, each oriented film turned out to have a single preferential switching: 71° switching in (001)$_{pc}$ oriented films, 109° switching in (110)$_{pc}$-oriented films and 180° switching in (111)$_{pc}$ oriented films, as determined below.

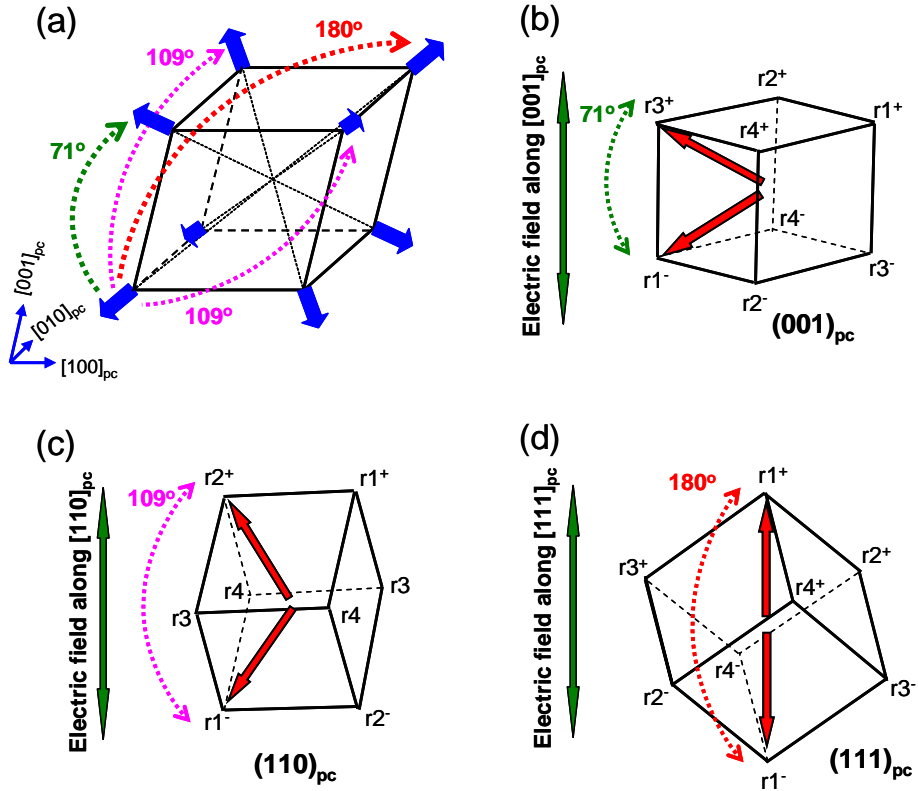

**Figure S1.** a) Possible polarization switching path. Schematic of a BiFeO$_3$ unit cell of different orientations and their switching paths: b) 71° switching in (001)$_{pc}$-, c) 109° switching in (110)$_{pc}$- and d) 180° switching in (111)$_{pc}$-oriented BiFeO$_3$ films.

**Determination of polarization switching path**

Figure S2a shows a reciprocal space map (RSM) around the 113 peak of the SrTiO$_3$ substrate and the as-grown (001)$_{pc}$ BiFeO$_3$ film exhibiting an initial monodomain state (r1). Using electrical measurements and piezo force microscopy (PFM), the polarization direction was determined to point downward, hence r1$^-$. We note that all three orientations of BiFeO$_3$ films in this work have a downward polarization direction, attributed to the preferential screening effect of the SrRuO$_3$ bottom electrode. After polarization switching occurred by applying a negative electric bias to 200 μm × 700 μm Pt top electrode arrays covering ~30% - ~40% area of the sample, the RSM in figure S2b shows an additional peak corresponding to an r3 domain, indicating 71° polarization switching from r1$^-$ to r3$^+$. The r3 domain disappeared after positive electric bias was applied to the top electrode, in figure S2c. Hence, 71° switching path is reversible by the vertical electric field in (001)$_{pc}$ BiFeO$_3$ films, and no other switching paths were observed. The same experiments were performed to determine the exact switching path in (110)$_{pc}$- and (111)$_{pc}$-oriented monodomain BiFeO$_3$ films.

Figure S2d shows the RSM of the as-grown (110)$_{pc}$ monodomain BiFeO$_3$ thin film near the 222 peak of the SrTiO$_3$ substrate. The figure shows a single 222$_{pc}$ peak of BiFeO$_3$ corresponding to r1 ferroelastic domains, which indicates a single r1$^-$ ferroelectric domain due to preferential downward polarization. After polarization switching by a negative electric bias was applied to the 200 μm × 700 μm Pt top electrode arrays, the RSM exhibited an additional 222$_{pc}$ peak corresponding to r2 domains, figure S2e, indicating that the polarization was switched from r1$^-$ to r2$^+$ by 109°. When the polarization was switched back by a positive electric field on the top electrodes, the r2 diffraction peak disappeared, as shown in figure S2f. Hence, 109° switching path is reversible by the vertical electric field in (001)$_{pc}$ BiFeO$_3$ films, and no other switching paths were observed.

Figure S2g shows the RSM of the as-grown (111)$_{pc}$ monodomain BiFeO$_3$ thin film near the 113 peak of the SrTiO$_3$ substrate. The RSM of the as-grown (111)$_{pc}$ oriented BiFeO$_3$ shows a single 113$_{pc}$ peak corresponding to the r1 ferroelastic domain, hence the r1$^-$ ferroelectric domain. In contrast to the (001)$_{pc}$- and (110)$_{pc}$-oriented BiFeO$_3$ films, no additional ferroelastic domain formation was observed after the polarization was switched up and down by a vertical electric field on 200 μm × 700 μm Pt top electrode arrays, as shown in figure S2h and S2i. Hence, 180° switching path is reversible by the vertical electric field in (001)$_{pc}$ BiFeO$_3$ films, and no other switching paths were observed. The measured $P_r$ value of the (111)$_{pc}$ BiFeO$_3$ film is consistent with the ratio of the geometric projection of spontaneous polarization in other orientations: $P_r(001_{pc}) : P_r(110_{pc}) : P_r(111_{pc}) = 67\,\mu C\,cm^{-2} : 92\,\mu C\,cm^{-2} : 115\,\mu C\,cm^{-2} \approx 1 : \sqrt{2} : \sqrt{3}$. This clearly indicates that the (111)$_{pc}$ BiFeO$_3$ films have 180° polarization reversal.

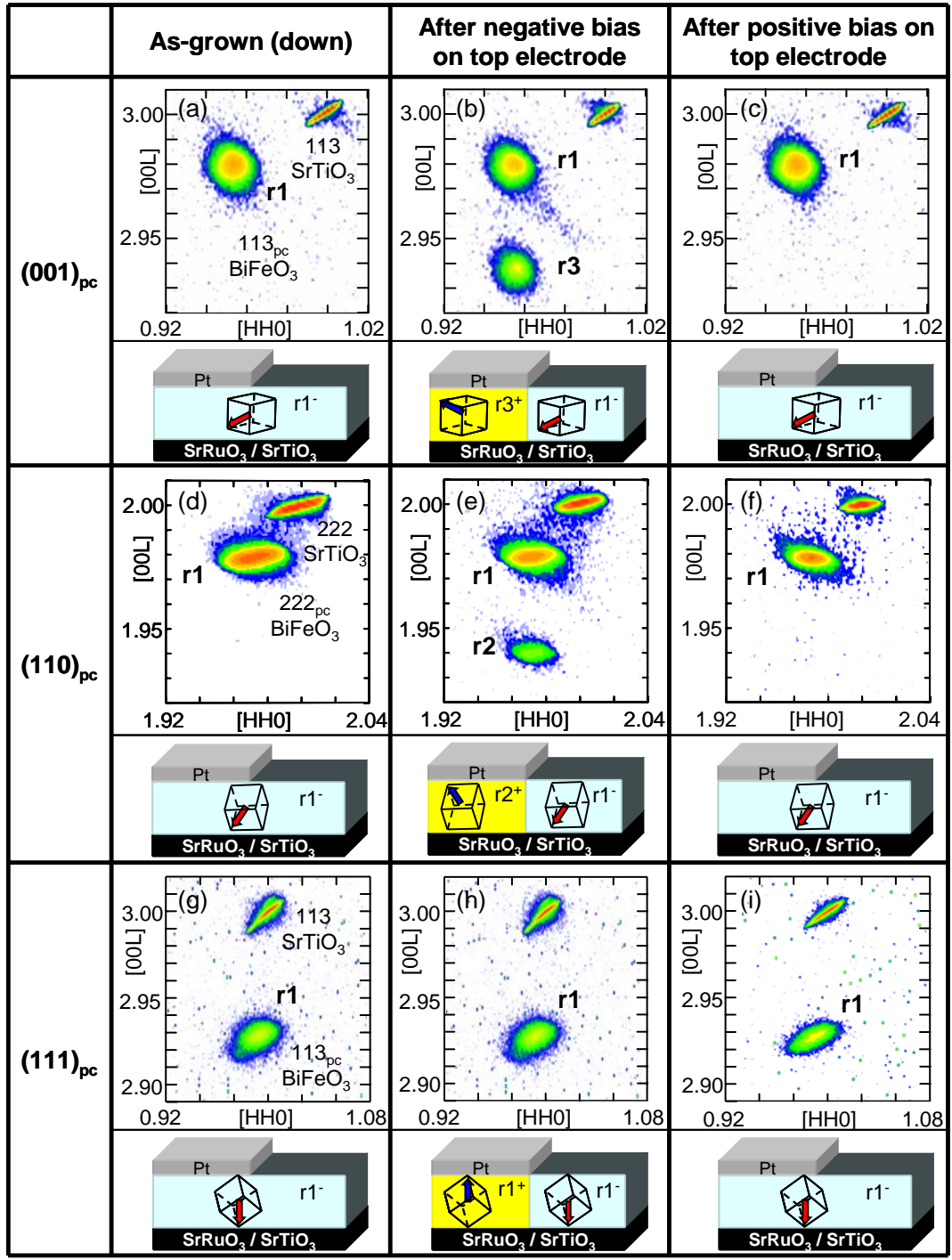

**Figure S2.** Reciprocal space mapping of a), b), c) (001)$_{pc}$-oriented and d), e), f) (110)$_{pc}$-oriented, and g), h) and i) (111)$_{pc}$-oriented BiFeO$_3$ films. The first column a), d), g) is for the as-grown monodomain state. The second column b), e), h) is for the state after switching to the up state. The third column c), f), i) is for the state after switching back to the down state. Schematic diagrams representing the domain structures of each RSM are shown at the bottom. It should be noted that all the BiFeO$_3$ samples have smooth SrRuO$_3$ bottom electrodes even though RSM data do not show SrRuO$_3$ peaks due to the absence of the diffraction peak corresponding to the 113$_{pc}$ and 222$_{pc}$ in the orthorhombic SrRuO$_3$ layer.

**PFM tip alignments and possible contrasts**

PFM images in figure 2 were collected with the alignment between PFM tip and BiFeO$_3$ film as shown in figure S3. The possible contrasts and corresponding domains are listed in the below.

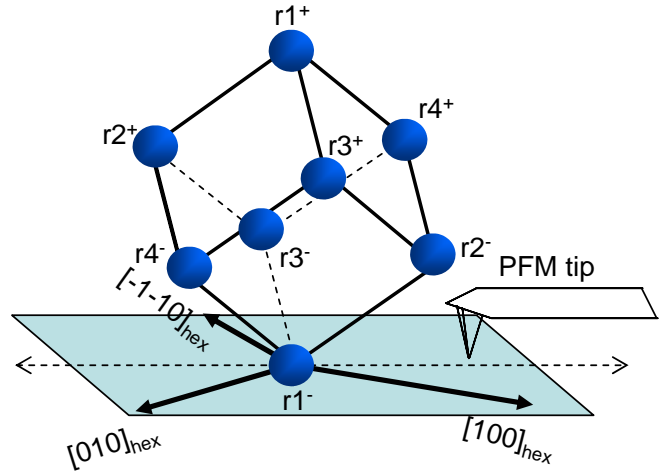

| OP image (4-levels) | IP image (3-levels) | Domain |
|---|---|---|
| Dark | Light | r1$^+$ |
| Dim-light | Dark | r2$^+$ or r4$^+$ |
| Dim-light | White | r3$^+$ |
| Bright-light | Dark | r3$^-$ |
| Bright-light | White | r2$^-$ or r4$^-$ |
| White | Light | r1$^-$ |

**Figure S3.** Schematic of PFM tip alignment with respect to the (111)$_{pc}$ BiFeO$_3$ film and possible contrasts corresponding to ferroelectric domains. The OP image has a maximum of 4 contrasts in a sequence of brightness: dark < dim-light < bright-

light < white. The IP image has a maximum of 3 contrasts in a sequence of brightness: dark < light < white.

**RSM data of (001)$_{pc}$ and (110)$_{pc}$ BFO after fatigue cycles**

Figure S4 a and b show RSM data of (001)$_{pc}$ BFO films before and after ~10$^6$ switching cycles, respectively. The initial monodomain state of as-grown BFO film does not change with switching cycles. Likewise, (110)$_{pc}$ BFO films exhibit a monodomain state after ~10$^6$ switching cycles (figure S4 c and d). These results support that the polarization fatigue is closely related to the new ferroelastic domain formation.

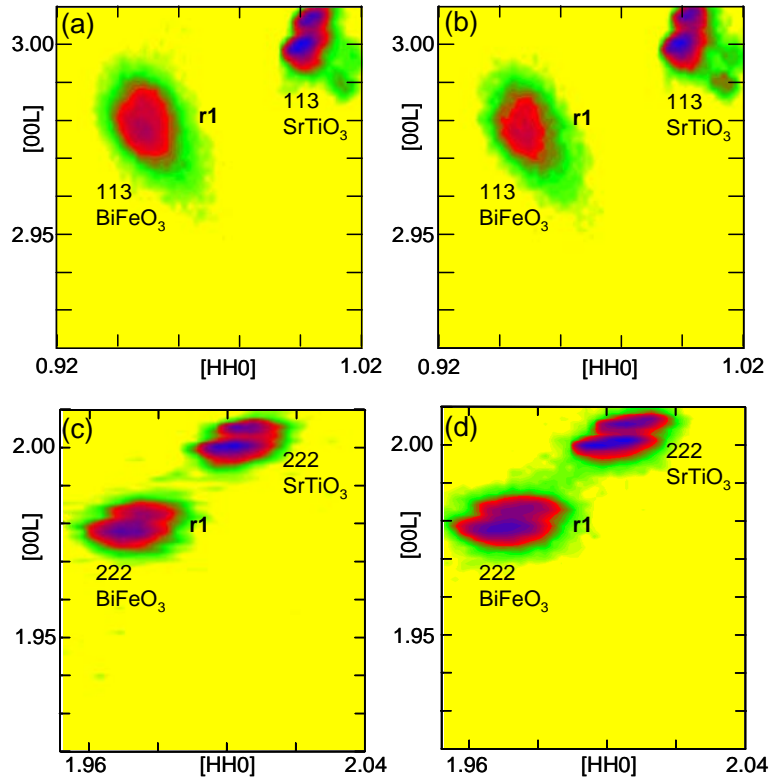

**Figure S4.** RSM data of BFO films before and after ~10$^6$ switching cycles. a) As-grown and b) after ~10$^6$ cycles of (001)$_{pc}$ BFO film. c) As-grown and d) after ~10$^6$ cycles of (110)$_{pc}$ BFO film.